\documentclass[a4paper,twocolumn,11pt,accepted=2026-05-10]{quantumarticle}
\pdfoutput=1
\usepackage{graphicx}
\usepackage{amsmath,amsthm,amssymb,dsfont,soul}
\usepackage{mdframed}
\usepackage{orcidlink}
\newcommand{\ket}[1]{\left| #1 \right\rangle}
\newcommand{\bra}[1]{\left\langle #1 \right|}

\newcommand{\braket}[2]{\langle #1|#2 \rangle}
\newcommand{\ketbra}[2]{\left|#1\right\rangle\hskip-1mm\left\langle#2\right|}

\usepackage{hyperref}
\usepackage[capitalise,nameinlink]{cleveref}

\begin{document}

\title{External quantum fluctuations select measurement contexts}

\author{Jonte R. Hance\,\orcidlink{0000-0001-8587-7618}}
\email{jonte.hance@newcastle.ac.uk}
\affiliation{Quantum Group, School of Computing, Newcastle University, 1 Science Square, Newcastle upon Tyne, NE4 5TG, UK}
\affiliation{Quantum Engineering Technology Laboratories, Department of Electrical and Electronic Engineering, University of Bristol, Woodland Road, Bristol, BS8 1US, UK}
\thanks{JRH acknowledges support from a Royal Society Research Grant (RG/R1/251590), an EPSRC Mathematical Sciences Small Grant (UKRI3647), and from their EPSRC Quantum Technologies Career Acceleration Fellowship (UKRI1217). The authors thanks John H. Selby for useful discussions.}
\author{Ming Ji\,\orcidlink{0000-0002-6569-5099}}
\affiliation{QICI Quantum Information and Computation Initiative, School of Computing and Data Science, The University of Hong Kong, Pokfulam Road, Hong Kong SAR, China}
\thanks{MJ acknowledges support from the National Natural Science Foundation of China via the Excellent Young Scientists Fund (Hong Kong and Macau) Project 12322516.}
\author{Tomonori Matsushita\,\orcidlink{0000-0001-7713-5374}}
\affiliation{Graduate School of Advanced Science and Engineering, Hiroshima University, Kagamiyama 1-3-1, Higashi Hiroshima 739-8530, Japan}
\thanks{TM was supported by JST, the establishment of university fellowships towards the creation of science technology innovation, Grant Number JPMJFS2129.}
\author{Holger F. Hofmann\,\orcidlink{0000-0001-5649-9718}}
\email{hofmann@hiroshima-u.ac.jp}
\affiliation{Graduate School of Advanced Science and Engineering, Hiroshima University, Kagamiyama 1-3-1, Higashi Hiroshima 739-8530, Japan}
\thanks{HFH was supported by ERATO, Japan Science and Technology Agency (JPMJER2402).}

\begin{abstract}
Quantum paradoxes show that the outcomes of different quantum measurements cannot be described by a single measurement-independent reality. Any theoretical description of a quantum measurement implies the selection of a specific measurement context. Here, we investigate generalised quantum measurements, in order to identify the mechanism by which this specific context is selected. We show that external quantum fluctuations, represented by the initial state of the measurement apparatus, play an essential role in the selection of the context. This has the non-trivial consequence that, when considering measurements other than just idealised projection-valued measures, different outcomes of a single measurement setup can represent different measurement contexts. We further show this result underpins recent claims that contextuality can occur in scenarios without measurement incompatibility.
\end{abstract}

\maketitle

\section{Introduction}

Contextuality, or the impossibility of assigning values to observables for a system such that those observables seem to describe measurement-independent properties of that system, is often viewed as one of the hallmarks of truly quantum systems, which cannot be described classically \cite{Kochen1968,Budroni2022KSContext}. Therefore, understanding the boundaries of contextual behaviour is key for us to be able to learn more about the classical-quantum divide.

In quantum mechanics, physical properties are described by self-adjoint operators, and the eigenstates and eigenvalues of a given operator are identified with the outcomes we get when we measure the value of its corresponding property (the property's measurement outcomes). As a result, it is not possible to define joint measurement outcomes for operators that do not share the same eigenstates. Instead, it is necessary to distinguish between different measurement contexts, defined by the specific set of eigenstates and eigenvalues relevant to a given measurement. Kochen and Specker formalised this definition of measurement contexts, and showed that the statistics observed in different measurement contexts can contradict each other. Cases where this contradiction occurs are referred to in this framework as \textit{contextual} \cite{Kochen1968,Budroni2022KSContext}.

A more recent definition of contextuality, given by Spekkens \cite{Spekkens2005Contextuality}, considers a scenario noncontextual only if it can be represented by a noncontextual ontological model (a form of hidden-variable model), where the probabilities of all measurement outcomes can be traced back to a probability distribution, over hypothetical states $\lambda$, which does not depend on the choice of measurement context. This identification of an underlying probability distribution for all possible measurements links contextuality with the failure to find a positive-valued probability distribution that consistently reproduces the measurement probabilities for all possible contexts, making it easier to connect quasiprobability negativity \cite{Spekkens2008NegContext} and through them, anomalous weak values \cite{Pusey2014AnomWVsContext}, to this ``generalised'' notion of quantum contextuality.

It is interesting to reflect on the differences between the mathematical approach to contextuality initially pioneered by Kochen and Specker, and the more recent approach introduced by Spekkens. Kochen and Specker mapped the problem of measurement contexts onto the eigenvalue spectrum of operators, effectively identifying the statistics of measurements with the probabilities obtained from an axiomatic application of Born's rule. On the other hand, Spekkens (through summations over the behaviours of ontic states represented by the same quantum state) refers to the probabilities obtained in actual measurements. This means that, in Spekkens' more general approach, the mathematical basis for the definition of a measurement context is less clear.

In a recent paper, Selby \textit{et al} pointed out that measurement incompatibility is neither necessary nor sufficient for a violation of Spekkens's generalised notion of noncontextuality~\cite{Selby2023CwoutI}. At first glance, this result may appear surprising, since it clearly undermines the idea that each set of compatible measurements defines a specific context. However, the result is less surprising when we consider that the original definition of contextuality relied on the identification of measurement outcomes with the eigenstates of self-adjoint operators, and so this approach always attributes {those} measurement outcomes {which we could describe using} non-orthogonal states to different measurement contexts~\cite{Ji2024quantitative}. This identification of contexts with state vectors in Hilbert space remains true even when these measurement outcomes, {despite corresponding to non-orthogonal state,} are obtained in the same measurement setup. Selby \textit{et al}'s result {(or at least one construction they give of it)} is based on the observation that different measurement contexts can be unified into a single measurement by constructing a positive operator valued measure (POVM), where the positive operators that represent different measurement outcomes need not commute with one another. Significantly, {for this construction,} Selby \textit{et al} suggest that the measurement outcomes can still be identified with the {values} of an underlying physical property, allowing them to effectively re-scale the experimentally observed probabilities to obtain modified conditions for the observation of contextuality. This appears to be at odds with the claim that the observed statistics cannot be explained by an underlying ontological model: certainly, the measurement outcomes of a POVM all have positive probabilities, and an ontological model could simply assign one possible reality to each outcome. It is therefore an essential assumption of Selby \textit{et al}\footnote{{At least when their framework is rendered in the language of quantum mechanics, like in~\cite{Selby2023CwoutI}, rather than in the language of GPTs, like in~\cite{Selby2023AccessibleFragments}.}} that it is reasonable to identify each measurement outcome of a POVM with the corresponding outcome of an ideal measurement projection onto an eigenstate, allowing them to re-scale the probabilities accordingly. 

Selby \textit{et al} base their arguments on the notion of ``operational equivalence,'' whereby measurement outcomes are considered ``equivalent''\footnote{{Up to some factor, which can then be removed by the rescaling process mentioned above, which is part of the ``flag convexification'' procedure they use to build a POVM from PVMs.}} whenever their probabilities are related to each other by a constant factor. However, it is left unclear {how this notion of operational equivalence relates to tangible experimental scenarios where we may be measuring such a POVM}. We aim to address this problem by considering the physics described by POVMs. In particular, we propose replacing the abstract mathematical notion of ``operational equivalence'' with an explanation of the physics by which a quantum measurement determines the context of each outcome. As we will show, measurement contexts can depend on quantum fluctuations in the environment, indicating that the choice of context in these scenarios is itself a quantum mechanical phenomenon.

{By quantum fluctuations, we here mean the randomness in the state of the (immediate) environment when an effective measurement is done to pick out one/a superposition of environmental states - e.g., by a coupling to an even larger observed environment. Given the possibility that these couplings to larger environments can effectively act like sequential measurements in non-commuting bases, the ``values'' of these quantum fluctuations are impossible to predict deterministically in advance (as with e.g., the energy fluctuations of the quantum vacuum, caused in at least some sense by finite-time measurements not commuting with energy measurements).} An important consequence of this result is, {given this unpredictability of quantum fluctuations,} contexts can be correlated with individual outcomes. This makes it impossible to identify contexts with a complete set of orthogonal basis states of Hilbert space. The selection of contexts by quantum fluctuations in the environment thus contradicts the assumptions of contextual ontological models that make the probability of a measurement outcome depend on the alternative outcomes that might have been observed instead of it. 

This paper is laid out as follows.
In Section~\ref{sec:conofqm}, we show that the original formulation of contextuality through the Kochen-Specker theorem is built on the idea of performing precise measurements with outcomes represented by orthogonal eigenstates (a projector-valued measure, or PVM), despite many realistic measurements not achieving this idealised level of precision, and so requiring a more realistic description of quantum measurements through POVMs. Since the elements of such POVMs needn't be orthogonal or normalised, two different outcomes of the same POVM can correspond to eigenstates of two non-commuting observables, and so, by an extension of the Kochen-Specker definition, be in two different measurement contexts. This challenges our intuition that a reproducible physical apparatus should always keep us within the same context, which motivates us to look for a better idea of what ``being in the same context'' means. We show that such a definition needs to be based on a clear separation of the ``object'' we are interested in and the environment, given we are interested in the physics of finite objects, rather than that of the Universe as a whole. We therefore argue that our notion of a context should be described entirely in terms of its effect on the system. 

In Section~\ref{sec:contextenv}, we consider how a specific context is established. To do this, we first describe measurement interactions between the system and its environment. The joint context of environment and system is described by entangled states which are orthogonal to each other. We show that one can obtain a POVM by applying the initial state of the environment to the joint context, defining a system context conditioned by the quantum mechanics of the environment. POVMs thus describe how quantum fluctuations in the environment define an objective measurement context.

In Section~\ref{sec:threepath}, we illustrate this selection of contexts using Hofmann's three-path interferometer \cite{hofmann2023sequential}. We introduce photon polarisation as a well-defined environmental degree of freedom, and derive different POVMs for the path degree of freedom based on a polarisation measurement performed after path and polarisation have been entangled by the insertion of a half wave plate in one of the paths. We show that this simple procedure results in the observation of outcomes from different measurement contexts, confirming our analysis in Section~\ref{sec:contextenv}.

In Section~\ref{sec:Relns}, we discuss the re-scaling procedure needed to demonstrate contextuality in a single measurement. We point out that the context selection in POVMs limits the maximal observable probability of a given outcome, providing an experimentally motivated alternative to Selby {\it et al}'s concept of ``operational equivalence''.

In Section~\ref{sec:cfd}, we discuss the implications of the environmental selection of a measurement context described by general POVMs. We point out that the measurement context is not defined by a complete set of possible measurement outcomes, making it impossible to identify the measurement context with a set of counterfactual results that were not actually obtained when a given outcome is observed. It is therefore important to revise the notion that measurement contexts refer to a specific orthogonal basis of Hilbert space, meaning we should change our perspective on the original formulation of contextuality as a relation between self-adjoint operators.

Finally, in Section~\ref{sec:disc}, we summarise our argument, and discuss its consequences.

{Note, in this paper we do not refer to or use eigenvalues, so all measurement outcomes are defined by their eigenstates (or, for POVMs, the eigenstate of each of the elements of the measurement), and we use the term measurement outcomes to refer to these. This is as Kochen and Specker define a context by a set of self-adjoint operators that have a non-degenerate eigenstate system. Therefore, a complete measurement context is characterized by measurement outcomes represented by PVMs of rank 1, and it is sufficient to refer to these eigenstates as (representations of) a measurement outcome. For POVMs, the logic is the same – a complete definition of context requires that the elements of the POVM are of rank 1. Note that we also do not consider subsequent measurements (we use the Born rule only), and never discuss post-measurement states - therefore, no collapse model is needed.}

\section{Contextuality of quantum measurements}\label{sec:conofqm}
In any theory of physics, it is crucial that we can relate the mathematical elements of the theory to observable properties of a physical system. In classical physics, this relation is quite trivial---what you (can) see is what you get---at least in principle.\footnote{This even applies to areas of classical physics which are typically not viewed as ``fundamental'': the microscopic degrees of freedom in statistical mechanics are in principle observable. On the other hand, thermodynamics only refers to macroscopic observations, even where abstract concepts such as entropy or free energy are concerned. For both, every mathematical expression is rooted in experimentally verifiable facts. Contextuality however messes with this expectation.}
Unfortunately, quantum mechanics throws a metaphorical wrench into this expected relation by separating our description of states from the description of possible measurement outcomes, leaving only the statistics given by Born's rule to mediate the two. Born's rule itself describes only precise measurements of physical properties, identifying measurement outcomes with the eigenstates of the self-adjoint operator that represents that observable of a quantum system in the Hilbert space formalism. It follows from this mathematical representation that it is not possible to construct a single measurement that determines the precise values of two observables represented by non-commuting operators. So long as observables $\hat{A}$ and $\hat{B}$ don't commute, either a physical setup measures the observable $\hat{A}$, or it measures the observable $\hat{B}$. The two measurements are incompatible and so must be represented by distinct physical situations. Since it is the physical setup that decides which measurement is performed, it is convenient to identify the choice of the measurement basis as the ``measurement context''. This definition of context entered the literature through the Kochen-Specker theorem \cite{Kochen1968}, and continues to provide the only consistent reference point for discussions of quantum contextuality.

However, realistic measurements are not precise. This is most obvious when thinking about position and momentum. In our usual experience of moving objects, we necessarily observe both position and momentum simultaneously. However, the precision of both observations is very far from the quantum mechanical limit set by the uncertainty principle. In general, it is always possible to perform joint measurements of non-commuting observables by choosing a particular uncertainty trade-off \cite{Englert1996Inequality,Matsushita2021Uncertainty}. Such quantum measurements are described by positive operator valued measures (POVMs), which assign a probability to each outcome according to a seemingly-straightforward generalisation of Born's rule.
Notably, the elements of a POVM need not be orthogonal or normalised, so long as they sum to identity. As a result, two different outcomes of the same POVM can correspond to eigenstates of two non-commuting observables, suggesting that the same measurement can provide precise information about either of these two non-commuting observables, depending on the outcome of the measurement.

A simple example of this would be the SIC-POVM for a qubit, which has four elements, corresponding to the four different measurement outcomes $\ket{0}$, and $(\ket{0} + \sqrt{2}e^{2ni\pi/3} \ket{1})/\sqrt{3}$, where $n$ is 0, 1, or 2. $\ket{0}$ is an eigenstate of $\sigma_z$, {so the projector on this state is an element of the PVM defined by $\sigma_z$}. Taking the other measurement outcome of the qubit SIC-POVM which also has real amplitudes in this basis representation, $(\ket{0} + \sqrt{2} \ket{1})/\sqrt{3}$, we see this is also a measurement outcome of the PVM defined by $(\sqrt{2}\sigma_x-\sigma_z)/\sqrt{3}$ ({this operator having} eigenstates $(\ket{0}+\sqrt{2}\ket{1})/\sqrt{3}$ and $\frac{1}{\sqrt{3}}(\sqrt{2}\ket{0} - \ket{1})$, which correspond to measurement outcomes of this PVM). These two PVMs do not commute: their commutator is $-2i\sqrt{2}\sigma_y/\sqrt{3}$. This shows how two different outcomes of the same POVM can correspond to eigenstates of two non-commuting observables.

It is a key question in quantum foundations whether repeatedly implementing a reproducible physical apparatus/protocol means that you always stay within a single context---while it may seem like the answer should be yes, we need to consider whether the decision to switch between contexts may in some circumstances be made within the apparatus itself. It is therefore important to have a clear idea what the essential concept of ``being in the same context'' actually is. 

Since we are interested in the physics of a finite object, our definition of a context needs to be based on a clear separation between the physical properties of the object, and the physics of the environment, which naturally includes any external apparatus that controls the interactions with the system. It is important that the context should be described entirely in terms of its effects on the system, without any explicit characterisation of the environment itself. As will be shown in the following, each element of a POVM is an ``objectified'' description of quantum correlations between the system and the environment, which can introduce a quantum-mechanical dependence of the measurement context on the precise outcome of the measurement. 

\section{Context selection by the environment}\label{sec:contextenv}

The physics of POVMs is more difficult to understand than the physics of precise measurements, since it is non-trivially determined by the interaction between the system and its environment. If we extend the Hilbert space to include the environment, different (overall, system-plus-environment) measurement outcomes must correspond to orthogonal states in this extended Hilbert space, since the outcomes are macroscopically distinct. This is true even if on a subspace, the different measurement outcomes {are represented by non-orthogonal state vectors}, and so the measurement is a POVM rather than a PVM (i.e., any POVM locally is really a PVM globally, over the combined system and environment---a physical interpretation of Neumark's Dilation Theorem \cite{gelfand1943imbedding}). 
In the product space of system and environment, the outcomes can thus be written as an orthogonal basis $\{\ket{m}_{ES}\}$ of combined states of the environment $E$ and the system $S$. {Here, we always assume that the measurement is maximally resolved with respect to the system, so that the global PVM can be represented by rank 1 projections. The PVM defined by the orthogonal basis $\{\ket{m}_{ES}\}$} represents a joint context of the system and its environment, but this joint context includes correlations between the system and the environment, which makes it impossible to identify a separate context of the system alone. {This means that the objective meaning of a specific measurement result $\{\ket{m}_{ES}\}$ depends on the initial state of the environment.} {It is therefore necessary to} include the initial state of the environment $\ket{\varphi_{\mathrm{init}}}_E$ as an additional constraint {that describes the effect of the environment on the measurement outcome represented by the global measurement result $\{\ket{m}_{ES}\}$.} 
Since this initial constraint changes the meaning of the measurement outcomes {with respect to the physics of the system}, it must be included in the measurement information {obtained from the system}. This is done by applying an inner product in the subspace of the Hilbert space that describes the environment,
\begin{equation}
    \ket{\lambda(m)}_S =\; _{E}\braket{\varphi_{\mathrm{init}}}{m}_{ES}.
\end{equation}
{Here, the state $\{\ket{m}_{ES}\}$ is an entangled state (over the system and environment Hilbert spaces) that describes the (accidental) correlations responsible for the result $m$, while the quantum fluctuations of the environment described by $\ket{\varphi_{\mathrm{init}}}_E$ assign quantum amplitudes to the contribution that originates from the environment only. The result is a description of the objective causes of the measurement outcome, i.e. the quantum state of the system that can be associated with the outcome once the quantum statistics of the environment have been accounted for.}
The additional information provided by the initial quantum state of the environment changes the orthonormal states describing collective physical properties of the system and the environment into a set of non-normalised non-orthogonal states $\{\ket{\lambda(m)}_S\}$ {(over the Hilbert space of the system alone)} which describe the properties of the system derived by applying the correlations of the joint state $\{\ket{m}_{ES}\}$ to the known properties of the environment $\ket{\varphi_{\mathrm{init}}}_E$. 

To better understand the role of the environment, we should first consider how this formalism describes the projection on an orthogonal basis $\{\ket{a}_S\}$ representing the eigenstates of an observable $\hat{A}$. In this case, we should obtain only a small subset of outcomes $\{\ket{m=a}_{ES}\}$, and each of these outcomes can be written as a product state,
\begin{equation}
    \ket{m=a}_{ES} = \ket{\varphi_{\mathrm{init}}}_E \otimes \ket{a}_{S}.
\end{equation}
Note that the other outcomes $m$ all have probability zero since they are orthogonal to the initial state of the environment.
{In a realistic scenario, this represents the experimental efforts necessary to suppress the effects of external noise sources. For instance, a large number of experiments depend on various cooling and filtering methods to ensure that the immediate environment of the system can be described by a pure state $\ket{\varphi_{\mathrm{init}}}_E$. In these cases, the appearance of $\ket{\varphi_{\mathrm{init}}}_E$ in the measurement outcome $\ket{m=a}_{ES}$ represents the necessary experimental confirmation that the cooling or filtering processes have worked. This is the sense in which the environment selects the context when an experiment is tailored to respond to a single orthogonal basis of the system.} What if the context depends on fluctuations in the initial state? We can illustrate this possibility with a simple extension of the projective measurement above. The measurement outcome $m$ now describes the selection of a context $x$ represented by a unitary transformation $\hat{U}_x$ and the subsequent outcome $a$,
\begin{equation}
\label{eq:switch}
    \Big|m=(x,a)\Big\rangle_{ES} = \ket{x}_E \otimes \hat{U}_x^\dagger \ket{a}_{S}.
\end{equation}
This results in a POVM described by the non-normalised states
\begin{equation}
\label{eq:switchPOVM}
    \ket{\lambda(x,a)} = \braket{\varphi_{\mathrm{init}}}{x} \hat{U}_x^\dagger \ket{a}_{S}.
\end{equation}
Here, the initial state of the environment describes quantum fluctuations in $\ket{x}_E$, which select a random context by controlling the quantum dynamics of the system. {In an experimental setup, $\hat{U}_x$ represents the measurement setting - the rotation angle of a polarizer, for instance. If such settings are realized by a quantum coherent interactions where the environmental degree of freedom described by the eigenstates $\ket{x}_E$ controls the setting of the measurement performed on the system, then quantum fluctuations select the measurement context in a manner similar to a classical random input. Outside of such an operational experimentally-motivated perspective, we could instead consider the environment as the uncontrollable larger ``bath'' of noise, which all we can do is mitigate against. In this sense, while the unitaries $\hat{U}_x$ act only on the system Hilbert space, they are still controlled by (i.e., which one is applied is determined by) the state of the environment, hence the label $x$ corresponding to the environmental state $\ket{x}_E$. For this understanding of the environment, the shift in the system caused by the unitary is determined by the (in this case random) state of the environment when an effective measurement is done on the environment to pick out one/a superposition of different $\ket{x}_E$. As we discuss below, this effect has some similarity to Zurek's notion of einselection~\cite{Zurek2003Einselection}.} 

This example shows {how it is possible that} the context selected in a single measurement setup {depends} on quantum fluctuations in the environment. {Note there will still be a non-trivial effect on the system from the fluctuations $\hat{U}_x$ even if the environmental state $\ket{\varphi_\text{init}}$ isn't a single $\ket{x}_E$, but a superposition of $\{\ket{x}_E\}$ (so long as we aren't in the special case where the weighted effect of the superposition of $\{\hat{U}_x\}$ applied to the system is equivalent to Identity).} But what about the general case, where it is not possible to identify a complete orthogonal basis set $\{\hat{U}_x^\dagger \ket{a}_{S}\}$? To understand this case, we need to consider possible correlations between the selection of an outcome in the system and the fluctuations of the environment. Since environmental fluctuations can introduce measurement errors, it is not possible to factorise the quantum fluctuation in $E$ and the measurement outcome in $S$. Instead, the joint measurement outcomes $\ket{m}_{ES}$ can be written as
\begin{equation}
\ket{m}_{ES} = \ket{\varphi_{\mathrm{init}}}_E \otimes \ket{\lambda(m)}_{S} + \ket{\sigma(m)}_{ES},
\end{equation}
where the additional states $\ket{\sigma(m)}_{ES}$ represent the part of the joint outcome that is orthogonal to the initial conditions. The non-orthogonal states $\ket{\lambda(m)}_{S}$ of the POVM are now included in the formulation, since they depend on the specific form of the quantum fluctuations in the initial state. The remaining state components $\ket{\sigma(m)}_{ES}$ can have any form, but it must ensure the states $\ket{m}_{ES}$ are mutually orthogonal through
\begin{equation}
\braket{\sigma(m)}{\sigma(m^\prime)} = - \braket{\lambda(m)}{\lambda(m^\prime)}, 
\end{equation}
and that the states $\ket{m}_{ES}$ are normalised through
\begin{equation}
\braket{\sigma(m)}{\sigma(m)} + \braket{\lambda(m)}{\lambda(m)} = 1.
\end{equation}
POVMs thus correlate the measurement errors of an uncertainty-limited quantum measurement with the selection of measurement contexts by the same quantum fluctuation of the environment. 

An important consequence of the outcome-dependent selection of a context is that we cannot identify the measurement context with a complete set of alternative measurement outcomes. This contradicts the idea of a context as an orthogonal basis, and the related idea that a context is determined by a set of commuting operators which can be measured jointly. The original definition of contexts based on observables seemingly requires counterfactual assumptions: what other outcomes could have been obtained in addition to the one actually observed. If the context is correlated with the outcome, such a counterfactual definition of context is not possible. Outcomes cannot depend on hypothetical alternative results, since these alternative results may be associated with different contexts. Once we generalise measurements to POVMs, we must define contextuality through the relation between specific outcomes, where two distinguishable outcomes share a context if and only if they are represented by orthogonal states, indicating that they represent alternative values of the same physical property. 

\section{Selection of contexts in a three-path interferometer}\label{sec:threepath}

\begin{figure}
    \centering
    \includegraphics[width=\linewidth]{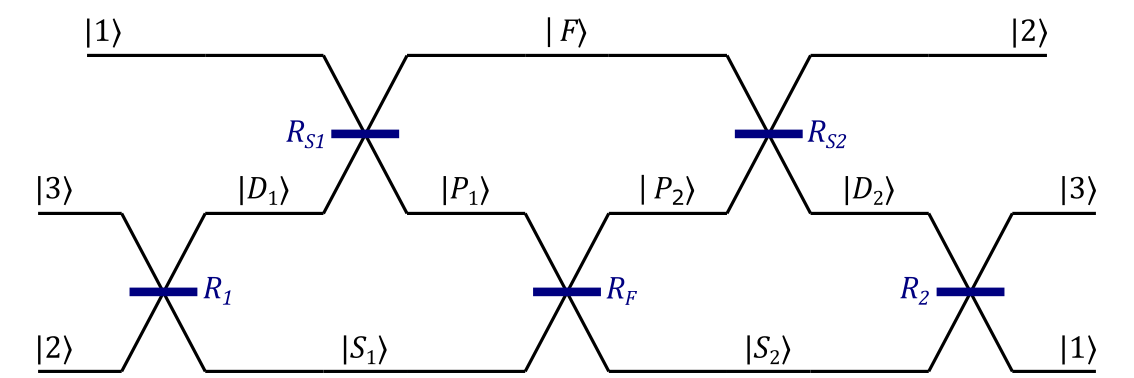}
    \caption{The three-path interferometer introduced by Hofmann to illustrate quantum contextuality \cite{hofmann2023sequential}. Each path represents a potential measurement outcome. The interferometer is aligned so that paths equivalent to the original input paths are reconstructed at the output, albeit with the position of paths $\ket{1}$ and $\ket{2}$ swapped. Relevant here is that path $\ket{F}$ is orthogonal to paths $\ket{S_1}$ and $\ket{S_2}$, but paths $\ket{S_1}$ and $\ket{S_2}$ are not orthogonal to one another.}
    \label{fig:threepath}
\end{figure}

We can illustrate this selection of contexts using the three-path interferometer recently introduced by Hofmann \cite{hofmann2023sequential}. In this interferometer, each path represents a potential measurement outcome that could be obtained by placing a detector into the path. Similarly to a Mach-Zehnder interferometer, the interferometer maps the input paths onto corresponding output paths, defining the orthogonal basis $\{\ket{1},\ket{2},\ket{3}\}$. The beam splitters represent unitary transformations of the basis states, defining the state $\ket{S1}$ ($\ket{S2}$) as an equal superposition of $\ket{2}$ and $\ket{3}$ ($\ket{1}$ and $\ket{3}$). Symmetry then requires that the central path $\ket{F}$ is an equal superposition of the input states,
\begin{equation}\label{eq:F}
    \ket{F}=\frac{1}{\sqrt{3}} \left(\ket{1}+\ket{2}-\ket{3} \right).
\end{equation}
If we used counterfactual assumptions to define the context of $\ket{F}$, we would need to distinguish between the outcome $\ket{F}$ obtained when $\ket{S1}$ was possible, and the outcome $\ket{F}$ obtained when $\ket{S2}$ was possible. As $\ket{F}$ is orthogonal to both $\ket{S_1}$ and $\ket{S_2}$, if we measure the particle as being on path $F$, we know it was not, nor will it be, on either paths $S_1$ or $S_2$: $\ket{F}$ is a joint eigenstate of the three projectors $\Pi_F$, $\Pi_{S1}$ and $\Pi_{S2}$, with eigenvalue +1 for $\Pi_F$, and eigenvalue 0 for $\Pi_{S1}$ and $\Pi_{S2}$. However, $\Pi_{S1}$ and $\Pi_{S2}$ do not commute, and therefore describe different measurement contexts. As our analysis above has shown, a POVM can select either of these contexts based on environmental quantum fluctuations. The detection of $\ket{F}$ must always exclude {\it both} $\ket{S_1}$ and $\ket{S_2}$, since a counterfactual identification of contexts is not possible. Therefore, in a POVM, it would be impossible to distinguish these two contexts.

It should be emphasised that this conclusion---that the observation of a specific measurement outcome cannot depend on any hypothetical alternative outcomes which were never actually observed---is strongly supported by the fact that the Hilbert space formalism defines each measurement outcome independently by a Hilbert space vector which provides no information about other possible outcomes. A specific projection can appear as an element in a wide variety of POVMs without ever changing its mathematical form. The only relation between different outcomes is the requirement that their probabilities must add up to one. 

For illustration, let us apply the general theory from the previous section to the three-path interferometer shown in Fig.~\ref{fig:threepath}. For this purpose, we introduce the polarisation of the photons in the interferometer as a source of environmental quantum noise. Let the vertical polarisation $\ket{V}$ pass through the interferometer unchanged, so that the output ports detect the input basis $\{\ket{1},\ket{2},\ket{3}\}$ for $V$ polarised outputs,
\begin{equation}
\Big|m=(V,i)\Big\rangle = \ket{V}_E\otimes \ket{i}_S,
\end{equation}
where $i=1,2,3$ is the output port in which the photon is detected. To change the context of the horizontal polarisation $\ket{H}$, a half wave plate is inserted in the path $\ket{F}$ described in Eq.~(\ref{eq:F}).
This phase change transforms the measurement context into
\begin{equation}
\begin{split}
    \Big|m=(H,1)\Big\rangle &= \ket{H} \otimes \frac{1}{3} \Big(\ket{1}-2\ket{2}+2\ket{3} \Big),\\
    \Big|m=(H,2)\Big\rangle &= \ket{H} \otimes \frac{1}{3}\Big(-2 \ket{1}+\ket{2}+2\ket{3} \Big),\\   
\Big|m=(H,3)\Big\rangle &= \ket{H} \otimes \frac{1}{3}\Big(2\ket{1}+2\ket{2}+\ket{3} \Big).
\end{split}
\end{equation}
If the initial polarisation state is diagonally polarised, so that 
\begin{equation}
    \ket{\varphi_{\mathrm{init}}} = \frac{1}{\sqrt{2}} \left(\ket{H}+\ket{V}\right),
\end{equation}
the POVM is described by the states
\begin{equation}
\begin{split}
\label{eq:Vcontext}
      \ket{\lambda(V,1)} &= \frac{1}{\sqrt{2}} \ket{1},\\
    \ket{\lambda(V,2)} &= \frac{1}{\sqrt{2}} \ket{2},\\   
\ket{\lambda(V,3)} &= \frac{1}{\sqrt{2}} \ket{3}
\end{split}
\end{equation}
for the selection of the original context by the detection of vertical polarisation, and
\begin{equation}
\begin{split}
\label{eq:Hcontext}
      \ket{\lambda(H,1)} &= \frac{1}{3\sqrt{2}} \left(\ket{1}-2\ket{2}+2\ket{3} \right),\\
    \ket{\lambda(H,2)} &= \frac{1}{3\sqrt{2}}\left(-2 \ket{1}+\ket{2}+2\ket{3} \right),\\ 
\ket{\lambda(H,3)} &= \frac{1}{3\sqrt{2}}\left(2\ket{1}+2\ket{2}+\ket{3} \right)
\end{split}
\end{equation}
for the selection of the transformed context by the detection of horizontal polarisation. This is an example of the situation characterised by Eqs.~(\ref{eq:switch}) and (\ref{eq:switchPOVM}). 

We can now modify the POVM by detecting a different set of polarisations in the output. Let us consider the diagonal and anti-diagonal polarisations
\begin{equation}
\begin{split}
    \ket{D}&=\frac{1}{\sqrt{2}} \left( \ket{H}+\ket{V}\right),\\
    \ket{A}&=\frac{1}{\sqrt{2}} \left( \ket{H}-\ket{V}\right).
\end{split}
\end{equation}
The corresponding measurement projections are described by entangled states of polarisation and path,
\begin{equation}
\begin{split}
    \ket{m(D,i)}&= \ket{H}\otimes \ket{\lambda(H,i)}+\ket{V}\otimes \ket{\lambda(V,i)},\\
    \ket{m(A,i)}&= \ket{H}\otimes \ket{\lambda(H,i)} {-}\ket{V}\otimes \ket{\lambda(V,i)}.
\end{split}
\end{equation}
{Note that these states are normalized, while the elements of the POVM $\ket{\lambda(H,i)}$ and $\ket{\lambda(V,i)}$ are not.}
For initial state $\ket{\varphi_{\mathrm{init}}} = \ket{D}$ (as before), the POVM now reads
\begin{equation}
\begin{split}
\label{eq:POVM2}
    \ket{\lambda(D,1)} &= \frac{1}{3} \left(2\ket{1}-\ket{2}+\ket{3} \right),\\
    \ket{\lambda(D,2)} &= \frac{1}{3}\left(- \ket{1}+2\ket{2}+\ket{3} \right),\\   
\ket{\lambda(D,3)} &= \frac{1}{3}\left(\ket{1}+\ket{2}+2\ket{3} \right)\\   
\ket{\lambda(A)} &= \frac{1}{\sqrt{3}}\left(\ket{1}+\ket{2}-\ket{3} \right),
\end{split}
\end{equation}
where the three results for antidiagonal polarisation can be summarised by a single state $\ket{\lambda(A)}=\ket{F}$ since the path outcomes are completely random and unrelated to the input state. Specifically, all three outcomes are given by
\begin{equation}
    \ket{\lambda(A,i)} = \pm \frac{1}{\sqrt{3}} \ket{F},
\end{equation}
where $i=1,2,3$. {It should be noted that the sign of the state has no observable effect, and cancels when the POVM is written in operator form.} If the path in which an antidiagonally-polarised photon is detected is ignored, the normalisation of the general outcome $A$ is brought back up to one, as indicated in Eq.~(\ref{eq:POVM2}). 

The three states $\ket{\lambda(D,i)}$ are non-orthogonal with inner products of
\begin{equation}
\begin{split}
    \braket{\lambda(D,1)}{\lambda(D,2)} &= -1/3,\\
    \braket{\lambda(D,2)}{\lambda(D,3)} &= 1/3,\\
    \braket{\lambda(D,3)}{\lambda(D,1)} &= 1/3.
\end{split}
\end{equation}
The detection of diagonal polarisation $\ket{D}$ seems to correspond to the input polarisation. However, the half-wave plate in path $\ket{F}$ changes the necessary input condition to anti-diagonal polarisation $\ket{A}$, so that the initial condition of $\ket{\varphi_{\mathrm{init}}} = \ket{D}$ removes the $\ket{F}$ component from the states $\ket{\lambda(D,i)}$ representing the detection of diagonal polarisation in the output. The components $\ket{\sigma(D,i)}$ of the joint states $\ket{m=(D,i)}$ are given by
\begin{equation}
\begin{split}
    \ket{\sigma(D,1)}&=\frac{1}{\sqrt{3}} \ket{A}\otimes\ket{F},\\
    \ket{\sigma(D,2)}&=\frac{1}{\sqrt{3}} \ket{A}\otimes\ket{F},\\
    \ket{\sigma(D,3)}&= - \frac{1}{\sqrt{3}} \ket{A}\otimes\ket{F}.
\end{split}
\end{equation}
The addition of these components associated with anti-diagonal input polarisations restores orthogonality to the joint path-polarisation measurements. 

Note that the two-dimensional Hilbert space of photon polarisation is sufficient to induce the selection of measurement contexts by environmental quantum fluctuations. Within the Hilbert space of the system (i.e., the paths of the interferometer), the outcomes now correspond to different contexts, where---similarly to $\ket{S1}$ and $\ket{S2}$---each of the three outcomes $\ket{\lambda(D,i)}$ shares a context with the outcome $\ket{\lambda(A)}=\ket{F}$ due to their orthogonality, making it impossible to distinguish between the contexts associated with the different outcomes $\ket{\lambda(D,i)}$ when the outcome $\ket{F}$ is obtained instead. The POVM described by Eq.~(\ref{eq:POVM2}) thus confirms that contextuality should be defined in terms of the relations between specific pairs of outcomes, where orthogonal pairs share a context and non-orthogonal pairs do not. Perhaps more generally, measurement outcomes share a context if and only if their projectors commute, independent of the other elements in the POVM that describes the complete measurement. 

\section{Relations between different contexts in the same measurement}\label{sec:Relns}

Selby \textit{et al} \cite{Selby2023CwoutI} argue that contextuality does not require incompatibility, because we can rescale the elements of a POVM to derive the probabilities that would be observed in a precise projective measurement of the corresponding observable. They refer to this possibility as ``simulability'', implying that POVMs {are ``operationally equivalent''} to precise projective measurements if their elements are proportional to each other. Their main example is a situation similar to the one described in Eq.~(\ref{eq:switchPOVM}) above, where the POVM describes the random selection of an orthogonal measurement basis.

{Let us first say something about the relationship between compatibility of measurements and commutativity of the operators representing those measurements formally. For both PVMs and POVMs, commutativity implies compatibility – but, unlike for PVMs, the inverse isn’t necessarily true for POVMs. Compatibility doesn’t necessarily imply commutativity~\cite{Heinosaari2010Nondisturbing,Heinosaari_2016}.
The definition of compatibility is trivial when only PVMs are considered. If PVMs are defined by eigenstates of operators, it is obvious that the PVMs of non-commuting operators are not compatible. Compatibility is interesting here because the elements of a given POVM do not necessarily commute with one another.}

As we have shown above, compatible measurements can represent different contexts. It is therefore wrong to assume that different measurement contexts must be represented by incompatible measurements. The specific measurement context of each measurement outcome depends on environmental fluctuations, and can change each time a measurement is performed. {Our analysis suggests that Selby's approach implies that all ``operationally equivalent'' measurement outcomes share the same context, regardless of whether that context was decided by deliberate experimental settings or by quantum fluctuations in the environment.}
If an experimentalist can choose between different incompatible measurements $x$, this choice can always be included in the description of the experiment by defining a probability distribution $P(x)$ for the choice of the measurement setting $x$ each time the experiment is run. If one considers the reduction in the output probabilities of measurement setting $x$ by a factor of $P(x)$ to be unproblematic, all incompatible measurements are ``simulable'' in the sense of \cite{Selby2023CwoutI}. Measurement incompatibility only refers to the actual probabilities of outcomes defined by a POVM, not to their class of operationally equivalent outcomes. 

Each outcome of a POVM has a maximal probability given by
\begin{equation}
    P_\lambda(m) = \bra{\lambda(m)} \hat{\rho} \ket{\lambda(m)} \leq \braket{\lambda(m)}{\lambda(m)}.
\end{equation}
The maximal probability is achieved if, and only if, the input state $\hat{\rho}$ of the system is given by a normalised version of the element of the POVM,
\begin{equation}
    \hat{\rho}_{m} = \frac{\ketbra{\lambda(m)}{\lambda(m)}}{\braket{\lambda(m)}{\lambda(m)}}.
\end{equation}
To obtain the outcome $m$, the environmental quantum fluctuations must select the corresponding context. We can interpret the maximal probability $\braket{\lambda(m)}{\lambda(m)}$ as the probability that the context of $\ket{\lambda(m)}$ is selected. For a general state $\hat{\rho}$, the probability of the outcome $m$ can then be rescaled to reflect the intrinsic probabilities of the physical property identified by the eigenstate $\ket{\lambda(m)}$. This allows us to reproduce the fundamental constraints that apply to probabilities of mutually exclusive outcomes, e.g., for the sum of the re-scaled probability of two outcomes represented by orthogonal states ($\braket{\lambda(m_1)}{\lambda(m_2)}=0$), 
\begin{equation}
    \frac{P_\lambda(m_1)}{\braket{\lambda(m_1)}{\lambda(m_1)}}+\frac{P_\lambda(m_2)}{\braket{\lambda(m_2)}{\lambda(m_2)}} \leq 1.
\end{equation}
Using such rescaled probabilities, we can now derive modified bounds for non-contextual relations between the elements of a single POVM. Most notable are outcomes with a probability of zero, since those outcomes can be used to exclude possibilities from the input state conditions. This can be used to construct Hardy-like contextuality paradoxes, e.g., from the relation between the outcome $\ket{F}$ given above and the outcomes $\ket{D_1}$ and $\ket{D_2}$ orthogonal to the outcomes $\ket{1}$ and $\ket{2}$, respectively. As quantum superpositions, these relations read
\begin{eqnarray}
    \ket{F} &=& \frac{1}{\sqrt{3}} \ket{1} + \sqrt{\frac{2}{3}} \ket{D_1}
    \nonumber \\
    &=& \frac{1}{\sqrt{3}} \ket{2} + \sqrt{\frac{2}{3}} \ket{D_2}.
\end{eqnarray}
Since $\braket{1}{2}=0$, in a noncontextual model, the state being $\ket{F}$ seemingly requires either $\ket{D_1}$ or $\ket{D_2}$ \cite{Ji2024quantitative}. However, the state with $P(D_1)=0$ and $P(D_2)=0$ has a non-vanishing probability of $P(F)=1/9$. A quantitative evaluation of contextuality needs to rely on the maximal probabilities of each outcome. In general, we can always find a POVM that includes the outcomes $\ket{\lambda(F)}$,  $\ket{\lambda(D_1)}$ and $\ket{\lambda(D_2)}$, so that quantum contextuality is indicated by
\begin{equation}
\label{eq:ineq}
    \frac{P_\lambda(F)}{\braket{\lambda(F)}{\lambda(F)}} >  \frac{P_\lambda(D_1)}{\braket{\lambda(D_1)}{\lambda(D_1)}}+ \frac{P_\lambda(D_2)}{\braket{\lambda(D_2)}{\lambda(D_2)}}.
\end{equation}
This inequality is sufficient if it can be certified that the states that maximise $P_\lambda(D_1)$ and $P_\lambda(D_2)$ respectively both achieve 2/3 of the maximal probability of $P_\lambda(F)$, while the states $\ket{1}$ and $\ket{2}$ achieve the remaining 1/3. 

The discussion above shows that the notion of ``operational equivalence'' used by Selby {\it et al} corresponds to the identification of the probability of context selection $\braket{\lambda(m)}{\lambda(m)}$. One might even argue that context selection is a representative example of ``operational equivalence''. In fact, the main example given in Eq.~(5) of \cite{Selby2023CwoutI} is explained in terms of context selection.
Eq.~(5) of the paper defines a POVM that measures one of $|T|$ different contexts with a probability of $1/|T|$,
\begin{equation}
    \tilde{E}_{y,t}:=\frac{1}{|T|}E_{y|t},
\end{equation}
where, for each context $t$, the subset $\{E_{y|t}\}$ describes the projection on the orthogonal basis of Hilbert space that defines that context. This form is an example of the context selection described by our Eq.~\eqref{eq:switchPOVM} and corresponds to the example given in Eqs.~\eqref{eq:Vcontext},\ref{eq:Hcontext}, where $|T|=2$, and the two ``settings'' $t$ correspond to the selection of $V$-polarisation or $H$-polarisation.  

The justification of the rescaling of probabilities in \cite{Selby2023CwoutI} relies on Spekkens' definition of ``generalised contextuality,'' which enforces linear constraints as a means of ensuring consistency between all possible measurements. It is important to consider the implications of such constraints, since they change the meaning of ``contextual'' in a fundamental manner. Our analysis of context selection provides an alternative derivation of these linear constraints, which may help to explain what the additional assumptions behind Spekkens' ``generalised contextuality'' really mean.
As shown above, the linear constraints allow Selby {\it et al} {through their flag convexification construction} to identify the outcomes of a POVM with the projectors $\{E_{y|t}\}$ onto the eigenstates of a specific observable. Our identification of a context selection probability $\braket{\lambda(m)}{\lambda(m)}$ shows that there is a physical explanation for this assumption, which can also be formulated as a factorisation into two independent probabilities.
It would seem that such an identification with an actual factorisation into probabilities originating from separate stages of two processes must always be possible if the linear constraints imposed by Spekkens' ``generalised contextuality'' are valid.
To avoid possible misunderstandings that could be caused by unnecessary mathematical abstractions, we should strive to explain these constraints in terms of the reproducible relations between measurement outcomes and physical properties in quantum mechanics. Our analysis achieves this goal by clearly identifying the probability of selecting a given context as the inner product $\braket{\lambda(m)}{\lambda(m)}$ for each individual measurement outcome. 

Finally, it seems important to ask whether the violation of the inequality in Eq.~(\ref{eq:ineq}) or Selby \textit{et al}'s violation of the inequality in Eq.~(13) of \cite{Selby2023CwoutI} can really be considered equivalent to the violation of an inequality obtained from incompatible measurements relating to non-commuting observables.
Note that joint measurability imposes unavoidable restrictions on the use of quantum correlations in practical applications, as explored in \cite{Renner2024CompNoise}. Also note that experimental efforts to characterise the violation of Bell's inequalities have used POVMs to jointly measure the statistics of all four measurement settings of a Bell-type scenario \cite{Matsuyama2023,Virzi2024Bell}. Matsuyama \textit{et al} explained how the measurement uncertainties of joint measurements prevent a direct observation of a Bell's inequality violation, but identified the Cirel'son bound as the limit imposed on Bell's correlations by the positivity of experimentally observed probabilities \cite{Matsuyama2023,Hof}. It is interesting to consider whether ``operational equivalence'' means that it would be legitimate to interpret these results as a direct demonstration of a Bell inequality violation based on a deconvolution of the statistical noise added by the measurement uncertainties, especially since this paper provides a detailed characterisation of the error statistics associated with the measurement uncertainties.

Virzi \textit{et al} have recently applied weak measurements to directly obtain the correlations responsible for a Bell's inequality violation in a single measurement \cite{Virzi2024Bell}. This work seems to pose an even greater challenge to Spekkens' ``generalised contextuality,'' since it suggests that ``operational equivalence'' should be applied to measurements that are so weak that they do not even disturb the initial state in a meaningful way. Effectively, Virzi \textit{et al} suggest that the violation of an inequality can be demonstrated without even selecting a specific measurement context. {While our result provides limits on the extent to which we can apply such operational equivalence, no equivalent limit exists for Spekkens' notion of contextuality. Therefore, while it has been argued that it is doubtful Virzi \textit{et al}'s result constitutes a proper Bell inequality violation \cite{Genovese2025CommentVirzi,Kupczynski2025CommentVirzi,Lambare2025CommentVirzi,Genovese2025ReplyCommentVirzi}, in Spekkens' framework Virzi \textit{et al}'s result is ``operationally equivalent'' to a Bell test, so their would be considered equivalent to experimental violations of Bell inequalities, despite its significant issues. This illustrates the problem with the concept of unbounded ``operational equivalence'' which Spekkens' notion of generalised contextuality employs, and illustrates why} it is important to first clarify what we mean by a measurement context, and how we relate joint measurability to the statistics of physical properties defined by their eigenstates, before we start making claims this bold.

\section{Counterfactual Definiteness}\label{sec:cfd}

One key reason for the confusion inherent in {at least the titular claim of} Selby \textit{et al}'s paper is that, as a community, we still need to work out what we mean by a ``measurement context''. Selby \textit{et al} {make use of} certain formal assumptions to relate their POVMs to precise projective measurements, implying a somewhat mysterious ``equivalence'' of the outcomes of a POVM with the eigenstates of an observable. As we have shown above, the essential indicator of different contexts is the non-orthogonality of the states that represent the different measurement outcomes. We therefore obtain a sufficiently compact and reliable definition of measurement contexts if we agree that measurement outcomes share the same context if and only if they are represented by orthogonal quantum states \cite{Ji2024quantitative}. Based on this definition, it is obvious that POVMs can include outcomes from many different contexts. In addition, we find that it is not necessary to associate a context with a complete orthogonal basis of Hilbert space, even though this might seem to be the case when one considers the original definition of contexts in terms of the eigenstates of operator observables. 

Based on the analysis presented above, we intend to correct the faulty idea that contextuality relates to a complete set of possible outcomes, i.e., that counterfactuals are needed to determine a context. Instead, we have shown that each actual outcome sufficiently determines its own context by identifying the dependence of the context on the entanglement of the system with environmental quantum fluctuations. Since the entanglement correlates the context with the specific measurement outcome, our result contradicts the assumption that the environment must be capable of selecting a complete set of orthogonal states in order to define the measurement context.

The assumption violated by the role of entanglement in the selection of a measurement context is that of counterfactual definiteness \cite{hance2019counterfactualrestrictions}. In a POVM, it is impossible to assign causal necessity to an outcome---to say that a certain outcome necessarily must have occurred, given a certain preparation and measurement procedure; and so that, by selecting a specific preparation procedure, we can cause a specific outcome. Specifically, the outcome 1 is only obtained when it coincides with the corresponding environmental fluctuation. It is therefore impossible to exclude the possibility of outcome 1 when a different outcome is obtained. The set of actual outcomes described by a POVM do not describe a set of alternative values of a physical property, so that the outcome 1 simply excludes all inputs orthogonal to that outcome, without specifying a unique set of counterfactual options. The use of POVMs thus indicates that the observation of the outcome 1 cannot depend on the available counterfactual alternatives, excluding possible explanations of contextuality that make the observation of 1 dependent on the unobserved outcomes of the measurement by which 1 is obtained. 

We can now compare our definition of context with the mathematical definition in the original formulation. In that formulation, a context is defined by combining observables with degenerate eigenspaces in such a way that the degeneracy is lifted. This means that each context is defined by a complete orthogonal basis, where the relation between different contexts is established by the degeneracy of the eigenspaces. As shown in \cite{Ji2024quantitative}, it is possible to describe this relation in terms of individual measurement outcomes by introducing intermediate contexts that share eigenstates with both of the two contexts linked by the degeneracy of the subspace. This admittedly makes the relation between these contexts more complicated, but it demonstrates that Kochen-Specker contextuality can be explained in terms of contexts defined by projections onto pure states. This becomes important when one realises that the same measurement outcome can be obtained by measuring completely different commuting operators. Consider for the moment that a non-degenerate eigenstate is defined by the eigenvalues $A_1$ and $B_1$ of $\hat{A}$ and $\hat{B}$. According to Kochen and Specker, obtaining $A_1+B_1$ and $A_1-B_1$ in a measurement of $\hat{A}+\hat{B}$ and $\hat{A}-\hat{B}$ would be a completely different context, despite the fact that any experimentalist would claim that the two measurements are equivalent ways of jointly obtaining both $\hat{A}$ and $\hat{B}$. It would seem that the problem arises only because counterfactual definiteness compels us to assume that the value $A_1$ might change depending on whether we actually measure $\hat{B}$, or rather a different observable $\hat{C}$ instead. If we understand that the context is defined by the non-degenerate eigenstates of the complete set of operators, it is obvious that the value $A_1$ does not change, whether the second measurement refers to $\hat{B}$ or to $\hat{C}$. Instead, the context is defined by the second measurement within the eigenspace of $\hat{A}$ associated with $A_1$. 

Interestingly, \cite{Ji2024quantitative} effectively shows that the two approaches to defining contexts yield the same results, even though they are based on different definitions of contextuality. The definition we present however does not require us to identify a complete set of eigenstates, which we consider as an advantage over the mathematical definition in the original formulation.

Does contextuality in its original formulation actually have anything to do with measurement, or is it just about idealised observables? While counterfactual definiteness is related to measurement, the argument that a result changes when we change which measurement was performed is stronger than we can make using the quantum formalism---to make such an argument we would need to develop a separate theory of what measurement is, which might be much too restrictive and so problematic. Arguments or logical structures built on being able to make such counterfactual statements typically break when combined with quantum mechanics. It is an active area of research as to just how much such arguments break down in quantum scenarios, and the extent to which this corresponds to unaccounted-for effects of measurement backaction (see e.g., \cite{Hance2024CFBAIG}).

To some degree, we share Selby \textit{et al}'s motivation in \cite{Selby2023CwoutI}: that the assumptions underpinning our ideas about ``what measurement is'' in the conventional formulation of contextuality are too restrictive. Selby \textit{et al} however only consider Spekkens' generalised contextuality. Instead, we should consider how best to alleviate this restrictiveness for Kochen-Specker contextuality too, given these assumptions are less relevant for Spekkens' notion of noncontextuality than they are for Kochen and Specker's. One key fact we wish to emphasise is, in contrast to these assumptions, noncommutativity is not incompatibility. If we restrict ourselves to only considering ideal projective measurements, noncommutativity and incompatibility coincide, which is why they are often confused (especially in Kochen-Specker contextuality, which only considers ideal projective measurements)---but this is not true in general for arbitrary measurements (such as POVMs). We therefore need to specify an actual problem, or actual physical scenario, as otherwise people may commit to definitions and formalisations which are not realistic. This consideration of actual physical scenarios is also necessary given that it is debatable whether the measurement outcomes we observe are independent of other things we are doing, even if they are spacelike-separated from the measurement event. Classically, it really doesn't make sense in terms of a time sequence of two measurements, that the second measurement should have an impact on the first. This is as, for spacelike-separated events, we can always pick a reference frame in which the apparent order of the events is reversed. In quantum mechanics, however, we can think about such spacelike-separated ``sequential'' measurements in two different ways: either by projecting on a product state by first projecting on one part of product, then projecting on the other; or by comparing outcomes, and showing that one outcome doesn't necessarily exclude another. In classical physics, certain pairs of measurement outcomes may be dichotomous (``either/or''). However, quantum mechanics allows us to put such dichotomous outcomes into a superposition. This changes how we need to consider the effects of such outcomes, given we can no longer as easily construct counterfactual arguments, where we ask ``what if'' we observed a measurement result dichotomous to the one we actually observed. The quantum formalism doesn't invite us to speculate about such ``what-ifs'', which makes considering concepts like counterfactual definiteness in quantum mechanics inherently difficult.

Further, as we go from PVMs to POVMs, we begin to see issues with the whole idea of ``the measurement we do here can affect the measurement we do there''---can such an idea be entertained if other outcomes can be completely arbitrary, rather than one of a set of orthogonal projectors? This links into the issue of whether scenarios being ``mathematically the same'' (or, as in Selby \textit{et al's} case, the same up to a rescaling factor) in quantum mechanics means such scenarios are also ``physically the same''---which of these notions of two scenarios being ``the same'' we choose to use heavily affects the meaning of compatibility.

This is also what we see in a (complex) interferometer when we measure one path, then choose which paths to interfere with which others---it seems inconceivable that our choice of which other paths to interfere affects whether the photon was in the path we measured. Such peculiar counterfactual shifts, or ``jumps'', seem based on mathematical assumptions with no good support in the real world. The real problem with such contextual behaviour is counterfactual definiteness---if a certain quantum fluctuation in the environment got you measurement result $A$, it is then pointless to speculate about the validity of an alternative outcome $B$. Similarly, with weak values, we see a change in the weak value of $B$ as we postselect on different $A$s. If we treat the POVM as a postselection, the value of an observable can be given by the different weak values determined by the contexts of the various outcomes, illustrating the change of context defined by non-commuting elements of the POVM in terms of the different values obtained for the same observable. 

\section{Discussion}\label{sec:disc}

The idea of contextuality originates from the Hilbert space formalism and needs to be firmly rooted in the specific non-classical relation between measurement outcomes described by it. Adapting an idea of contextuality that allows non-orthogonal outcomes to be part of the same context undermines a consistent generalization of (non)contextuality to measurement scenarios described by POVMs.
As shown above, these cases are described by frustrated destructive interference in the larger Hilbert space of system and environment. The orthogonality we observe over this larger Hilbert space requires destructive interferences between components associated with different states of the environment, which can then be frustrated by the initial condition of the quantum mechanical environment. It is therefore important that the initial conditions determined by the environment are not considered as secondary, given that they are needed to specify the context within the system of interest. The reduction of the larger Hilbert space of system and environment achieved by the initial conditions is essential for an objective description of the physics of a system in terms of the observable effects that the system has on its environment. Ancillary degrees of freedom should not be considered as part of the system, as doing so would make it impossible to discuss the physics of the system in a consistent manner. 

{It is worth noting that the environmental degrees of freedom discussed here are actually part of the experimental setup by which a quantum system is accessed. It has often been noted (especially in Zurek's theory of einselection~\cite{Zurek2003Einselection}) that a natural measurement basis might be selected by corresponding decoherence processes inherent in the quantum statistics of a macroscopic environment. In our analysis, such a process might be responsible for the selection of the global PVM described by $\ket{m}_{ES}$. The environment we are describing is limited to the immediate vicinity of the quantum object, where experimentalists can control the fluctuations to the point where the environment can effectively be described by a pure state $\ket{\varphi}_{\mathrm{init}}$. It is indeed interesting to consider how an experimentalist can establish control over the immediate environment of a quantum system, and perhaps we would have to concede that a pure-state description would not be adequate at that point. What the present discussion has shown is that the immediate environment includes quantum fluctuations, even if the control over the quantum system is optimal. Therefore, the selection of contexts is best characterized by quantum processes that take place well before decoherence separates distinct measurement outcomes at the macroscopic level.}

The complete physics of a specific system are described by its own Hilbert space and the purpose of a measurement is to determine the statistics that characterize physical properties within this Hilbert space. An informationally complete characterization of these statistics necessarily requires measurements relating to different measurement contexts. This relation between informational completeness and contextuality is essential for our understanding of quantum statistics, since it explains why there is an unavoidable amount of negativity in all statistical approaches such as quasiprobabilities or symmetric informationally-complete POVMs (SIC-POVMs), either in the characterization of the joint statistics of separate measurements, or in the reconstruction of quantum states from the positive statistics of a POVM \cite{Hofmann2020OpAlgebra}. Based on the above definition of measurement contexts, this negativity can now be identified with the ``quantumness'' in the relation between different measurement contexts. 

It may also be worth noting that the constraint used by Selby \textit{et al} in \cite{Selby2023CwoutI} is in some sense similar to von Neumann's argument against hidden variables \cite{von2018mathematical,golub2024retrospective}. We could reduce {the flag convexification construction of} Selby \textit{et al}'s argument to the fact that it is not possible to find a linear decomposition of pure states into positive parts. This means that it is impossible to separate a state $\ket{a}$ into ($a$ AND $b$) and ($a$ AND NOT $b$). Similarly, the assumption of linearity in Spekkens's generalised (non)contextuality maps directly onto an operator expansion of the sort given in \cite{Hofmann2020OpAlgebra}. Specifically, the condition of linearity seems to require that hidden variable theories can be represented by an orthogonal and complete operator basis. Since any such basis is provably non-positive, quantum theory is contextual. It is likely that the assumption of linearity corresponds to von Neumann's requirements of a hidden variable theory that does not replace quantum mechanics with something different \cite{von2018mathematical,golub2024retrospective}. In both cases, Bohmian trajectories are excluded because they predict non-linear relations between initial and final conditions, so that particles can arrive at a different place even though the linear time evolution of the quantum state arrived back at the initial state. 

To summarise, we have shown in this paper that, when considering generalised quantum measurements (i.e., POVMs rather than projective measurements), the choice of measurement itself does not solely establish the measurement context---external quantum fluctuations (only represented when we also consider the initial state of the measurement device) also play a role in selecting the measurement context. This means compatible measurements, or even different outcomes of a single measurement setup, can put the system into different measurement contexts. Given the reliance of both Kochen-Specker and Spekkens' formulations of contextuality on the identification of idealised measurements with a complete Hilbert space basis, our result that individual measurement outcomes sufficiently define their own context has significant consequences for our understanding of quantum contextuality.

\bibliographystyle{quantum}

\end{document}